\documentclass[conference]{IEEEtran}
\IEEEoverridecommandlockouts
\usepackage{cite}
\usepackage{amsmath,amssymb,amsfonts}
\usepackage{algorithmic}
\usepackage{graphicx}
\usepackage{textcomp}
\usepackage{xcolor}
\usepackage{multirow}
\usepackage{multicol}
\usepackage{subfigure}
\usepackage{hyperref}
\usepackage{dblfloatfix}
\def\BibTeX{{\rm B\kern-.05em{\sc i\kern-.025em b}\kern-.08em
    T\kern-.1667em\lower.7ex\hbox{E}\kern-.125emX}}
\begin{document}

{\LARGE This paper is a preprint (IEEE “accepted” status). IEEE Copyright notice: \text{\textcopyright} 2019 IEEE. Personal use of this material is permitted. Permission from IEEE
must be obtained for all other uses, in any current or future media, including
reprinting/republishing this material for advertising or promotional purposes,
creating new collective works, for resale or redistribution to servers or lists, or reuse of any copyrighted component of this work in other works. \par}

\title{A Novel method for IDC Prediction in Breast Cancer Histopathology images using Deep Residual Neural Networks}

\author{\IEEEauthorblockN{Chandra Churh Chatterjee}
\IEEEauthorblockA{\textit{Computer Science and Engineering Department} \\
\textit{Jalpaiguri Government Engineering College}\\
Jalpaiguri, India \\
ccc2025@cse.jgec.ac.in}
\and
\IEEEauthorblockN{Gopal Krishna}
\IEEEauthorblockA{\textit{Computer Science and Engineering Department} \\
\textit{Jalpaiguri Government Engineering College}\\
Jalpaiguri, India \\
gk1948@cse.jgec.ac.in
}
}

\maketitle

\begin{abstract}
Invasive ductal carcinoma (IDC), which is also sometimes known as the infiltrating ductal carcinoma, is the most regular form of breast cancer. It accounts to about \(80\%\) of all breast cancers. According to American Cancer Society \cite{us_cancer}, more than \(180,000\) women in the United States are diagnosed with invasive breast cancer each year. The survival rate associated with this form of cancer is about \(77\%\) to \(93\%\)  depending on the stage at which they are being diagnosed. The invasiveness and the frequency of the occurrence of these disease makes it one of the difficult cancers to be diagnosed. Our proposed methodology involves diagnosing the invasive ductal carcinoma with a deep residual convolution network to classify the IDC affected histopathological images from the normal images. The dataset for the purpose used is a benchmark dataset known as the Breast Histopathology Images \cite{idc_data}. The microscopic RGB images are converted into a seven channel image matrix, which are then fed to the network. The proposed model produces a \(99.29\%\) accurate approach towards prediction of IDC in the histopathology images with an AUROC score of \(0.9996\). Classification ability of the model is tested using standard performance metrics. The following methodology has been described in the next sections.
\end{abstract}


\section{Introduction}
The breast cancer is the worst and the most common form of cancer occurring among women compared to other forms of cancer. Invasive carcinoma in breast cancer is actually classified in two categories as follows:
\begin{itemize}
    \item Invasive ductal carcinoma (IDC)
    \item Invasive lobular carcinoma (ILC)
\end{itemize}
The Invasive lobular carcinoma is the rarest form of breast cancer accounting to only \(10\%\) of all invasive carcinomas \cite{health_line}, where as invasive ductal carcinoma is the most common type of breast cancer affecting women especially in the age range of \(45\) to \(60\) or above according to Cancer Treatment Centers of America \cite{health_line}. One of the major cause of breast cancer is damage caused to a cell's DNA \cite{national_bf}. The IDC form of the breast cancer gets its name from its invasive nature and its ability to spread to other parts affecting the whole body. The early detection of invasive ductal carcinoma is of paramount importance for its early treatment. The diagnosis of IDC in breast cancer requires very severe treatment including surgeries and radiation therapies. pathological diagnosis requires a microscope and manual studying of the different slides to classify them into cancer positive or negative, This process is very time consuming and conveys a lot of errors due to human cognitive limitations which can easily be tackled using computer vision techniques associated with studying histopathological images.
\begin{figure}[ht!]
    \centering
    \includegraphics[width=2.5cm]{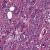}
    \includegraphics[width=2.5cm]{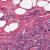}
    \includegraphics[width=2.5cm]{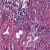}
    \caption{Invasive ductal carcinoma affected histopathological images}
    \label{fig:idc_img}
\end{figure}
\newline\indent
The proposed methodology is developed to aid the detection of IDC in breast cancer histopathology images. The dataset thus used for the purpose contained histopathological images which are very difficult to be analyzed using naked eye as shown in Fig.\ref{fig:idc_img}. The images have a very low resolution and, thus visualization of the following images is quite inconvenient. The image slides are whole slide images with a stain on them as present in most of the microscopic images. The proposed model is developed using the concepts of deep residual learning algorithm which allows better flow of the error throughout. To obtain the best performance results and to overcome the challenge of staining in histopathological images we merged 4 different image channels (hue, saturation channel of the HSV and \(l^*\), \(a^*\) channel of the CIELAB color space) with the RGB color space containing three channels thus resulting in images containing a 7 channel image matrix. The images have some amount of magnification on them resulting in the loss of resolution in the images that are captured from the microscope. The acquisition performed on these images results in generation of several artifacts which need to be removed for accurate detection of IDC. Image re-sizing is performed,which maintains the dimension and the aspect ratio of the core images and aids in the visualization of the images performed using GRAD-CAM. Noise removal is performed using Gaussian blurring algorithm which accurately removed the artifacts, maintaining the contrast ratio in the images. The validity of the proposed model was finally tested using GRAD-CAM visualization for the convolution layers and the dense or the fully connected (FC) layers.
\newline\indent
The rest of the paper describes the details of the working and evaluation of the model and is arranged in the following manner: Sect. \ref{sec:related_work} discusses the works related to IDC Prediction in Breast Cancer histopathology images using the existing statistical models and compares the performance of our model with them; Sect. \ref{sec:methodology} explains in details the dataset  preparation and model development phases; Sect. \ref{sec:result_discuss} discusses the results obtained by the model. Finally, Sect. \ref{sec:conclusion} concludes the study and discusses its future scope.

\section{Literature Survey}
\label{sec:related_work}

Many attempts have been made to automate the process of IDC detection in breast cancer histopathalogy images. The methods using machine learning and especially deep learning have been the most successful so far in progressing the task of carcinoma detection in histopathology images. This is because of the fact that deep learning methods like convolutional neural networks (CNN) \cite{lecun1998lenet} have become exceptionally well not only at image classification \cite{imagenet2012deep}, but also at all other kinds of visual recognition tasks like object-detection \cite{ren2015faster}, semantic segmentation \cite{he2017mask} etc. The digital analysis of microscopic images are explained by N. Dey et al \cite{dey}, where nuclei detection and segmentation are described, which helps in the understanding of histopathological images. K. He et al. proposed a novel type of CNN known as Residual Networks \cite{he2016deep} which further boosted the performance of CNNs by easing the process of training substantially deeper networks without the problem of vanishing gradients. Our proposed method also uses the concept of the mentioned residual networks.
\newline\indent
Several deep learning paradigms and data mining techniques are described by K. Lan et al \cite{lan} along with data preprocessing and transformation which is also an important part of bio-informatics. Image processing \& analysis along with machine learning methods have also been applied to classification of histopathology images. M. Veta et al. \cite{veta2014review} discusses image analysis methods for breast cancer histopathology images by tissue preparation using stains and image segmentation methods. S. Doyle et al. \cite{doyle2008grading} proposed a method of image analysis for distinguishing low and high grade breast cancer from histopathology images. S. Naik et al. \cite{naik2008grading} presented a glandular and nuclear segmentation method to classify prostate and breast cancer histopathology images. A level-set algorithm and a template matching algorithm in addition to a Bayesian classifier are used to utilize the pixel information at different scales. M. M. Dundar et al. \cite{dundar2011classification} proposed a setup to automatically classify the breast tissue images to differentiate between usual ductal hyperplasia   (UDH) and ductal carcinoma in situ (DCIS). B. E. Bejnordi et al. \cite{bejnordi2016detect} put forward an algorithm to automatically detect DCIS in digitized hematoxylin and eosin (H\&E) stained histopathological slides of breast tissue. This method utilizes multi-scale superpixel classification on whole slide images (WSIs) to distinguish DCIS from a large set of benign conditions. In \cite{bejnordi2017deep} B. E. Bejnordi et al. make use of convolutional neural networks for classification of hematoxylin and eosin (H\&E) stained breast specimens. Using this method they achieve an area under ROC of 0.92 for the given task. Similarly, M. S. Reza and J. Ma \cite{reza2018classify} employ different sampling techniques along with convolutional neural networks for histopathology image classification in class-imbalanced data. M. Balazsi et al. \cite{balazsi2016} proposed a system for detecting regions expressing IDC in images of microscopic tissues or whole digital slides. A. C. Roa et al. \cite{cruz2014cnn} proposed a CNN architecture for detection and analysis of IDC tissue regions in whole slide images (WSIs) of breast cancer. The WSIs were divided into non-overlapping image patches. A three layer CNN architecture is used. This consists of two layers of convolutional plus pooling layers, a fully-connected layer followed by a logistic regression classifier was used to predict if an image patch contains IDC tissue. B. E. Bejnordi et al. \cite{bejnordi2017stacked} recently proposed another methodology for classification of invasive ductal carcinomas in whole-slide histopathology images. The method proposes stacking two CNN on top of each other, where feature maps produced by the first CNN is fed into the second CNN. This system  achieves a three class accuracy of 81.3\% for classification of WSIs into normal/benign, DCIS, and IDC. It also achieves an AUC of 0.962 for the binary classification of non-malignant and malignant slides.

\section{Proposed Methodology}
\label{sec:methodology}
\begin{figure}[]
    \centering
    \includegraphics[width=.925\linewidth, height = 5cm]{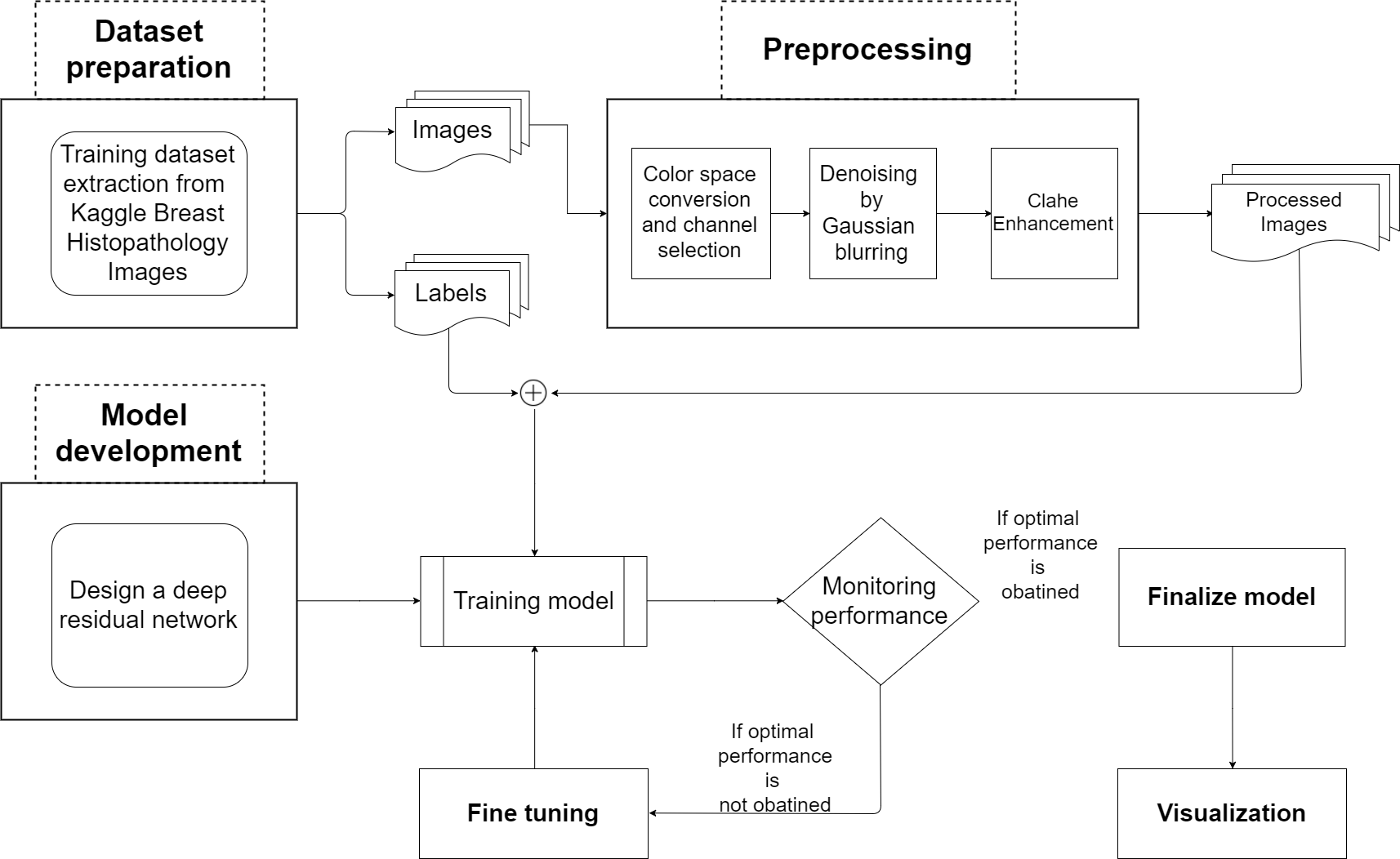}
    \caption{Overview of the proposed methodology}
    \label{fig:overview}
\end{figure}
\begin{figure*}[t!]
    \centering
    \includegraphics[width=.925\linewidth, height = 5cm]{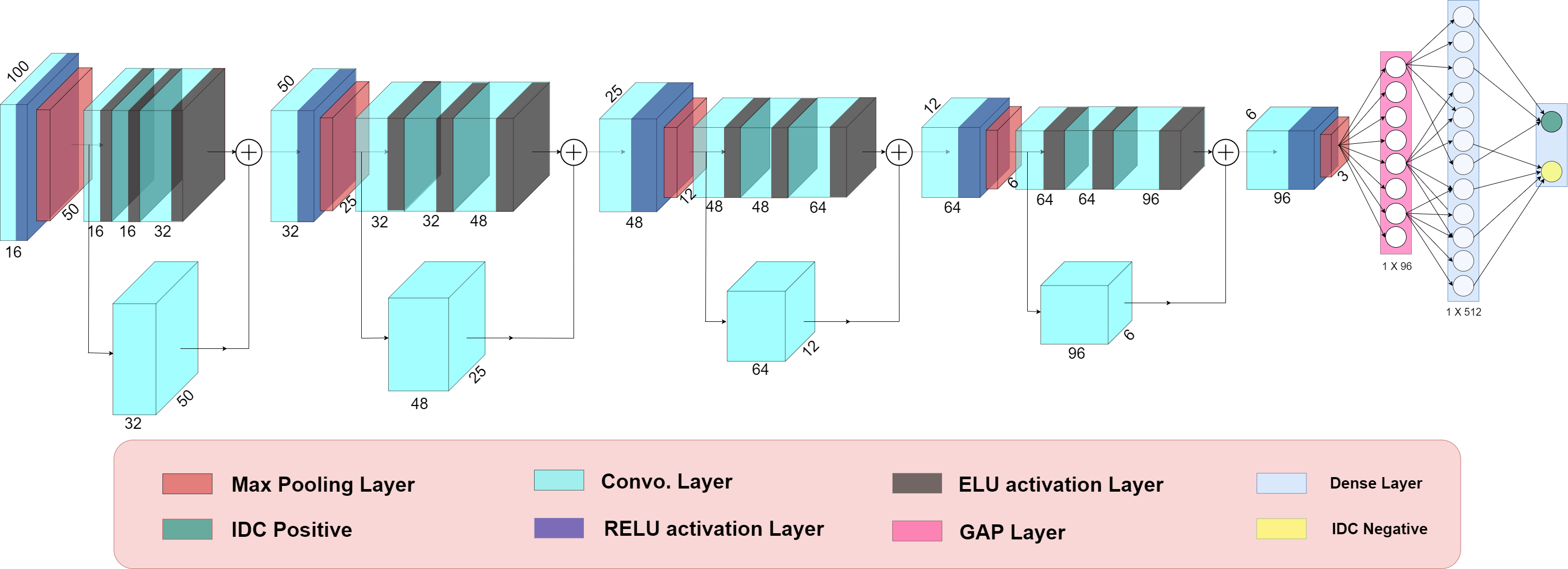}
    \caption{Model architecture}
    \label{fig:model}
\end{figure*}
Deep learning algorithms are very robust in nature while working with image data. The dataset in question consists of histopathological images of breast cancer affected cells and normal or unaffected cells. The proposed model revolves around the concept of deep residual learning using 2D-Convolutional layers. An extensive preprocessing done on the images involving channel extraction is proposed along with the model. The networks become prone to dangers of increasing loss(caused by the problem of \(vanishing gradient\)) depending on the depth of the model. Hence our proposed methodology involves paradigms to overcome such problems.
\newline\indent
Fig. \ref{fig:overview} illustrates the brief overview of the proposed methodology and Fig. \ref{fig:model}  discusses the 3D representation of the network. In the sections below the dataset preparation and the proposed methodology is discussed in details.

\subsection{Dataset Preparation}
The dataset was extracted from the Kaggle Breast Cancer Histopathology Images\cite{idc_data}. The dataset consists of 277,524 patches of size \((50 \times 50)\) which were originally extracted form 162 whole mount slide images of Breast cancer(BCa) specimens scanned at 40x. Out of all the images, 198,738 were diagnosed with IDC negative and 78,786 with IDC positive. The proposed methodology is developed on a subset of 7,500 images consisting of 3,000 IDC positive and 4,500 IDC neagtive images. A subset of images, rather than all the images was used to suffice for the problem of memory error.

\subsection{Data Preprocessing}
The images had a very less resolution of \((50 \times 50)\). Hence the images were resized to dimensions \((100  \times 100)\) to tackle the problem of low resolution of the images. The resizing was followed by selection of different channels or color spaces from the images such as (l* and a* channel from the CIELAB colorspace and hue and saturation channel from the HSV color space). The four different channels were merged with the original RGB colorspace to obtain a 7 channel image matrix to be fed to the model. The denoising was performed using the Gaussian blur algorithm followed by enhancement of the images performed using the CLAHE(contrast limited adaptive histogram equalization).

\subsubsection{Denoising of Images}
The denoising of the images was necessary due to low resolution which hampers the affected regions in the microscopic cells especially the IDC positive labelled images. The Gaussian blur algorithm was used for the denoising purpose. The Gaussian filter works based on a function defined as:
\begin{equation}
\label{eqn:denoise}
    G(x,y) = \frac{1}{2\pi\sigma^2}\times\exp{(-\frac{x^2 + y^2}{2\sigma^2})} .
\end{equation}
\newline\indent
From the Eqn. \ref{eqn:denoise}, it is clear that the function is separable in two dimensions, thus having less time complexity compared to other blurring algorithms. All the 7 channels undergo this blurring.

\subsubsection{Enhancement of the Images}
The images in the dataset are histopathological or microscopic in nature. Hence the images were enhanced using the CLAHE(contrast limited adaptive histogram equalization) algorithm to make the region of interest more prominent for better performance of the model. 
\newline\indent
Now the enhancement was performed only on the R, G and B channels of the RGB colorspace to prevent over-enhancement of the images. Over-enhancement can lead to completely different interpretation of the images by the model leading to wrong visualizations. It should be noted that the denoising was performed prior to enhancement, thus eliminating the probability of noise enhancement.
\newline\indent
In the CLAHE algorithm an image is represented as \(f(x,y)\) with histogram as \(h(i)\) and its cumulative distribution is defined as follows:
\begin{equation}
\label{eqn:cumu_hist}
    H(i) = \int_{0}^{i} h(p) dp
\end{equation}
The image is enhanced using the function defined as:
\begin{equation}
\label{eqn:enh_img}
    g(x,y) = H(f(x,y))
\end{equation}
Here, \(g(x,y)\) is the enhanced image which can be obtained for each individual channel of the image.

\subsection{Model Development}
The proposed model revolves around the concept of residual learning using 2D-convolutional layers. The 2D-convolutional layers are very efficient in dealing with image data. However with an increase in depth of the model the loss saturates at a  much higher value compared to a network with few layers. This effect of decreasing accuracy and saturating loss is known as \textit{vanishing Gradient}. The model architecture and a 3D illustration of the model is presented as Fig. \ref{fig:model}.

\subsubsection{Convolution Layer}
2D-convolutional layers were used for our model, which prove to be very efficient while dealing with images. The parameter complexity and the computation cost of a traditional convolutional algorithm is defined as:
\begin{equation}
\label{eqn:parameterComplexity}
    \rho = k \times k \times f_{in} \times f_{out}
\end{equation}
\begin{equation}
\label{eqn:computationalCost}
    \kappa = \rho \times d_{in} \times d_{in}
\end{equation}
The \(K\) represents the dimension of the kernel which in our case was considered to be \((4 \times 4)\). The \(f\) represents the feature maps provide input and obtained as output through each convolution. An image of resolution \((100 \times 100)\) was provided as input to the first convolutional layer. 

\subsubsection{Residual Module}
The proposed network incorporates the deep residual neural network paradigm. The residual module with the shortcut connection instills the residual learning approach. The increasing depth of a model results in the issue of degradation and an increasing loss. The solution to this problem is provided by Kaming He et al. \cite{he2016deep} who proposed the deep residual learning approach. Identity mappings or \textit{shortcut connections} are used to propagate the error through depth of the network as a result of which the error gradient propagates without any significant or considerable degradation. 
 \begin{figure}[h!]
    \centering
    \includegraphics[width=.7\linewidth, height = 4.5cm]{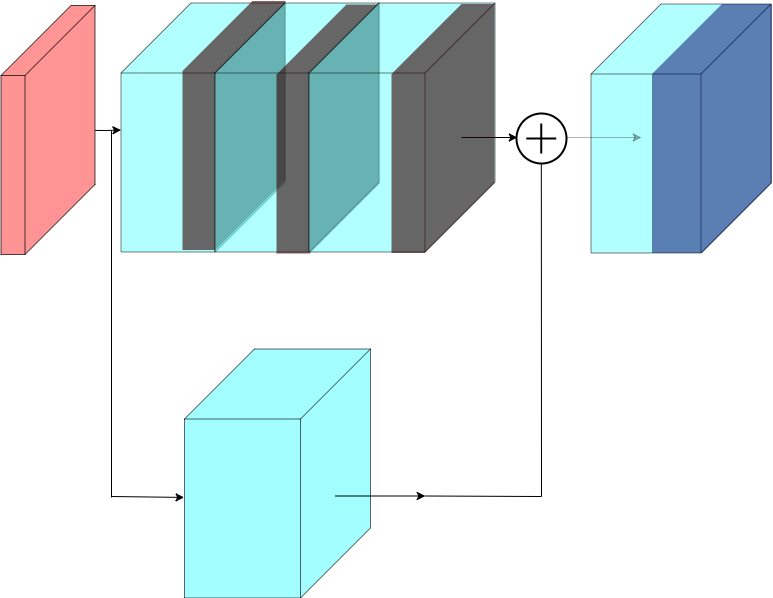}
    \caption{A single Residual Module from the proposed network}
    \label{fig:residualModule}
\end{figure}
 \newline\indent
 There are a total of \(4\) residual blocks in the proposed  model and each residual block consists of 3 convolutional layers and a residual connection with a single convolutional layer. The serial convolution layers are activated using the ELU activation where as no activation is provided to the convolutional layer in the shortcut connection. Addition is performed between the shortcut connection and the last convolutional layer of each residual block, followed by a RELU activation layer.
 \newline\indent
 Finally, a Global average pooling operation is performed to obtain a 1D-vector which is then fed to a FC(Fully connected layer). The residual module used in our study is illustrated in Fig. \ref{fig:residualModule}
 
\section{Results and Discussion}
\label{sec:result_discuss}
The proposed model is optimized based on monitoring the accuracy, loss, AUROC score and several other parameters such as Precision, Recall and \(F_1\) score. The model is also visualized using Grad-CAM(Gradient weighted class activation maps) to generate the heat map for better understanding of the region of interest.
\newline\indent
Fine tuning was performed by changing the kernel dimensions and the number of residual modules for the optimal performance of the neural network. The kernel dimension ranged from \((2 \times 2)\) to \((7 \times 7)\). The number of residual blocks ranged from 1 to 5 and not beyond due to huge increase in the parameter complexity of the model leading to large memory consumption. The model with residual blocks 1 to 5 except 4 are not used for comparison as they produced poor results as compared to the optimal model with 4 residual blocks. 

\subsection{Experimental Setup}
The proposed model architecture is built using Keras framework where RMSprop has been used as an optimizer which adjusts the learning rate automatically. The best weights of the model are saved using the early stopping technique. The model was trained on on a total of 6,000 images out of 7,500 and the rest was used for testing purpose. Out of the training set of images of 6,000 only 600 images were used for validation purpose. No images overlap with each other out of the training, testing and the validation set of images.

\subsection{Confusion Matrix and Performance scores}
\begin{figure}[h]
    \centering
    \includegraphics[width=.825\linewidth, height = 6.2cm]{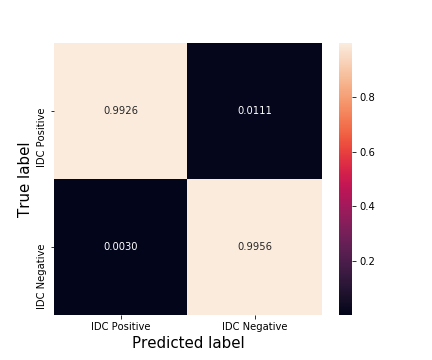}
    \caption{Confusion Matrix of the proposed network}
    \label{fig:conf_mat}
\end{figure}

The confusion matrix is illustrated in the Fig. \ref{fig:conf_mat}. It illustrates the performance of the proposed model on the test set. The confusion matrix is based on the model with 4 residual blocks consisting of convolutional layers with kernel dimensions of \((4 \times 4)\).
The performance metrics derived from the aforementioned confusion matrix is shown in the following Table. \ref{tab:perf_comp}.
\begin{table}[h!]
\caption{The performance scores of the proposed model}
\label{tab:perf_comp}
\begin{tabular}{|l|l|l|l|l|l|}
\hline
Class       & Accuracy & Precision & Recall & specificity & F1 Score \\ \hline
Affected & 0.9929   & 0.9969    & 0.9889 & 0.9969      & 0.99294  \\ \hline
Normal & 0.9929   & 0.9889    & 0.9969 & 0.9889      & 0.99296  \\ \hline
\end{tabular}
\end{table}
\begin{figure*}[!t]
    \centering
    \begin{subfigure}
        {\includegraphics[height=7cm, width=7.5cm]{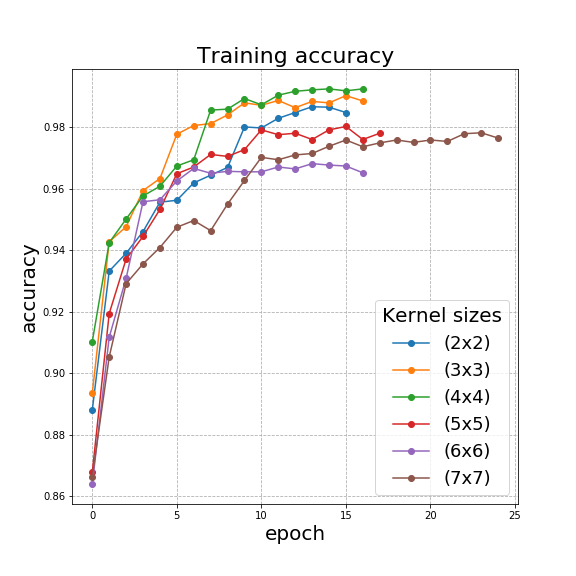}}
    \end{subfigure}
    \begin{subfigure}
        {\includegraphics[height=7cm, width=7.5cm]{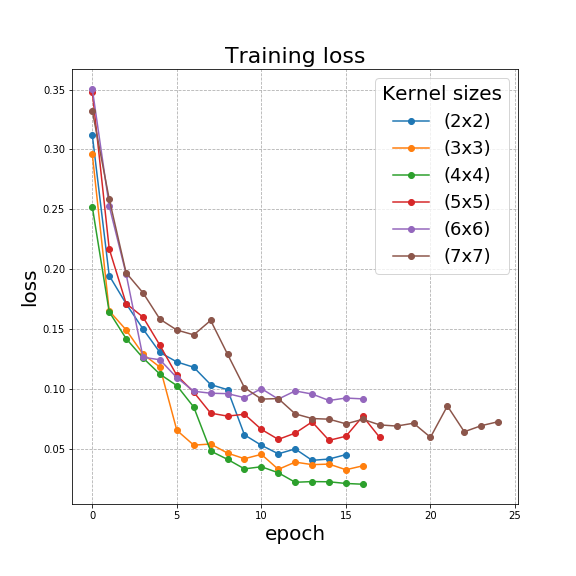}}
    \end{subfigure}
    \caption{The comparison of the training accuracy and training loss curves for different kernel dimensions for the network with 4 residual blocks} 
    \label{fig:performanceCurves}
\end{figure*}
\begin{figure*}[!b]
    \centering
    \includegraphics[width=3.5cm]{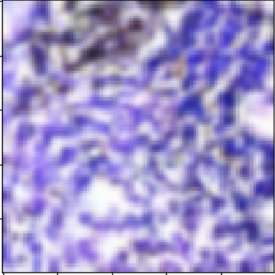}
    \includegraphics[width=3.5cm]{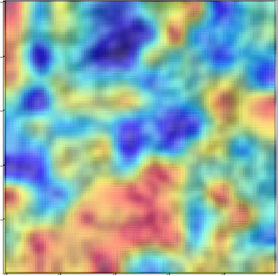}
    \includegraphics[width=3.5cm]{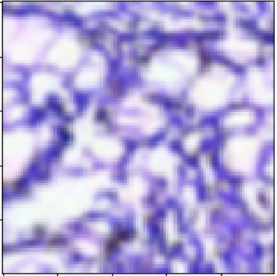}
    \includegraphics[width=3.5cm]{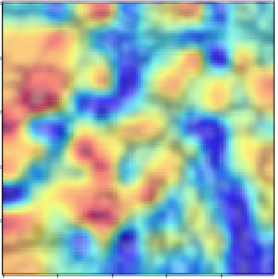}
    \caption{The Grad-CAM visualization of the images(Left: Original processed images; Right: Grad-CAM heat map of the ROI of the images)}
    \label{fig:grad_cam}
\end{figure*}
From the above table we can see that the model has achieved an over the top accuracy. The other performance metrics also validate the optimal performance of the proposed network and the methodology.
\begin{figure}[h]
    \centering
    \includegraphics[width=.825\linewidth, height = 6.2cm]{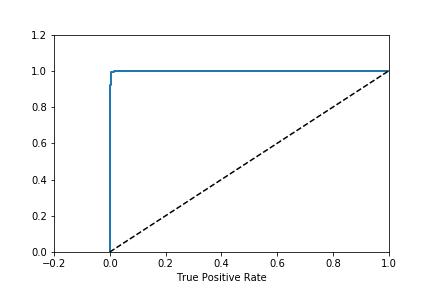}
    \caption{The ROC curve}
    \label{fig:roc}
\end{figure}
\subsection{Performance Curves}
The ROC curve of the model performance is displayed in Fig. \ref{fig:roc}. The AUROC score of the optimum model with 4 residual blocks, consisting of convolutional layers with kernel dimension of \((4 \times 4)\) is found to be \textbf{\(0.9996\)}. Hence the proposed model can be addressed as an one of a kind state-of-the-art-model.
The training accuracy and the loss curves for the model with 4 residual blocks consisting of convolutional layers with various kernel sizes ranging from \((2 \times 2)\) to \((7 \times 7)\) is illustrated in Fig. \ref{fig:performanceCurves}.
From the aforementioned figure it is clear that the model showed optimal performance with convolutional layers of kernel dimensions of \((4 \times 4)\).

\subsection{Grad-CAM Visualization}
The Gradient weighted class activation maps(Grad-CAM) for the individual convolutional layers were monitored to validate the classification ability of the model. The class activation maps help to visualize the affected regions in the histopathological images of the dataset. The heat map that is generated due to the Grad-CAM algorithm is a representation of the region of interest while preserving the spatial information, which is lost in the dense or the FC layers of the model.
\begin{table*}[!t]
\caption{Comparative analysis with the existing works}
\label{tab:comp}
\centering
\begin{tabular}{|l|l|l|l|}
\hline
Author & Year & Employed Methodology & Accuracy \\
&&& \\ \hline
M. S.  Reza et al. \cite{reza2018classify}& 2018 & Over sampling and under sampling over CNN based classifiers & 85.48\% \\
&&& \\ \hline
B. E. Bejnordi et al. \cite{bejnordi2017stacked}& 2017 & Stacked CNN for classification & 81.3\% \\ 
&&& \\ \hline
M. Balazsi et al. \cite{balazsi2016}& 2016 & Random Forest Classification over extensive feature extraction & 88.7\% \\ 
&&& \\ \hline
A. C. Roa et al. \cite{cruz2014cnn}& 2014 & CNN based classification on whole slide images & 84.23\% \\ 
&&& \\ \hline
\textbf{Proposed methodology} & -- & \textbf{Deep residual neural networks' based classification}  & \textbf{99.29\%} \\
&&& \\ \hline
\end{tabular}
\end{table*}
The Grad-CAMs of the last convolutional layer of the residual network for few images from the test set is illustrated in the Fig. \ref{fig:grad_cam} along with the original processed images.

\section{Conclusion and future scope}
\label{sec:conclusion}
A novel, state-of-the-art methodology for accurate diagnosis of IDC(Invasive Ductal Carcinoma) in breast cancer has been proposed in this study. Comparative study with the existing statistical models illustrated in Table. \ref{tab:comp} shows that the proposed model has better performance than the existing models for the classification of IDC in breast cancer histopathological images.
\newline\indent
The proposed methodology provides the importance to the different channels in the images that can be fed to any deep learning model. The different color spaces make the model more optimum in terms of classification ability of the model. The residual learning incorporated in the proposed methodology overcomes the problem of \textit{vanishing gradient} and accuracy degradation with increase in depth of the model.
\newline\indent
As a future scope of this study, it is intended to extend the model's classification ability towards diagnosing several diseases based on the corresponding histopathological images. It is also intended to make the proposed model a state-of-the-art algorithm for classification of different histopathological or microscopic images.

\bibliographystyle{IEEEtran}
\bibliography{Reference}

\end{document}